\documentclass[tikz,12pt]{article}
\usepackage{mheckII}

\usepackage[T1]{fontenc}

\usepackage{tikz-feynman}

\usepackage{manfnt}
\usepackage{longtable}

\usepackage{fancybox}

\usepackage{multirow}

\usepackage{blkarray}

\usepackage{tikz} 
\usetikzlibrary{matrix}

\theoremstyle{theorem}

\theoremstyle{definition}

\def\Dsl{\,\raise.15ex\hbox{/}\mkern-13.5mu D}
\def\dsl{\,\raise.25ex\hbox{/}\mkern-10.5mu \partial}

\title{Solving a puzzle in the rank 2 $\cn=2$ classification
by Argyres and Martone 
}

\authors{Sergio Cecotti\footnote{e-mail: {\tt cecotti@sissa.it}}\vskip 9pt

\centerline{SISSA, via Bonomea 265, I-34100 Trieste, ITALY}
}

\abstract{Argyres and Martone have produced a beautiful and deep classification of the scale invariant Special Geometries in rank 2.
They get a puzzle: the scale-invariant geometries with Coulomb dimensions $\{2,2\}$ appear to depend on \emph{four} free complex parameters,
while on physical grounds we expect \emph{only two} marginal deformations. We show that the \emph{isoclasses} of $\{2,2\}$
Special Geometries
are indeed parametrized by a complex space of dimension 2, in facts by a non-singular del Pezzo surface of degree 5,
a result which exactly matches the physical expectation by Gaiotto. This solves the puzzle.  
}

\begin{document}
\maketitle

\section{Introduction}

In \cite{Argyres:2022lah} Argyres and Martone have given a nice, deep, and explicit
classification of scale invariant Special Geometries in rank 2.
However they get a puzzling result. The geometries with Coulomb dimensions $\{2,2\}$
depend on \emph{four} complex free parameters (see eq.\eqref{curve} below)
while these geometries are expected to describe 
the  $SU(2) \times SU(2)$ superconformal gauge theory with a bi-fundamental hypermultiplet
and two fundamental hypermultiplets for each gauge factor, which has only \emph{two}
 exactly marginal deformations. Indeed this SCFT is the theory of class $\cs[A_1]$
 with Gaiotto curve \cite{Gaiotto:2009we} the sphere with five punctures, so its conformal manifold should be
 the moduli space of 5-punctured spheres, $\cm_{0,5}$ (whose canonical compactification is the
 del Pezzo surface of degree 5), which has complex dimension 2 not 4.
 
Indeed physics dictates the number of marginal deformations in a rank-$r$ 4d $\cn=2$ SCFT
 to be equal to the multiplicity of 2
 in the list of Coulomb dimensions, so finding a dimension 4 deformation space 
 for $\{2,2\}$ special geometries would be a kind of a counterexample.
 This would be quite surprising since this statement about the dimension of the conformal
 manifold can be proven directly in Special Geometry without any appeal to physical arguments.
 It seems that we have got a purely geometric paradox.
 
 In this short note we show that the space which parametrizes the \emph{inequivalent}
 $\{2,2\}$ special geometries constructed by Argyres and Martone in \cite{Argyres:2022lah} is indeed $\cm_{0,5}$
 exactly as predicted by Gaiotto.
 
 The geometric techniques we use are more interesting than the result itself.
 This is perhaps a justification for writing this otherwise ``obvious'' note.

\section{Review of scale invariant special geometry}
We review the relevant geometric facts mainly to fix the notation.
\medskip

A (rigid) special geometry is a holomorphic fibration $\pi\colon\cx\to \cc$, with section $s$, 
where $\cx$ is a holomorphic symplectic manifold, with holomorphic $(2,0)$ form $\Omega$,
while the smooth fibers of $\pi$ are Lagrangian submanifolds as well as polarized Abelian varieties.
The rank of the geometry is the complex dimension of the base $\cc$, called the \emph{Coulomb branch} (it is an affine variety).
The geometry is \emph{scale invariant} if, in addition, it admits an holomorphic Euler vector $\ce$ whose exponential is an automorphism for all the implied
geometric structures, i.e. for all $t\in\C$ the exponential $\exp(t \ce)\colon \cx\to\cx$
is an automorphism of the complex manifold $\cx$ which maps fibers into fibers and fixes (set-wise)
the section $s$,
 while
\be
\mathscr{L}_\ce\,\Omega=\Omega.
\ee 
A \emph{non-singular} scale invariant geometry has a base $\cc$ which is biholomorphic to
 $\C^r$ with coordinates $\{u_1,\cdots, u_r\}$
of definite scaling dimension $\Delta_i$\; \footnote{\ In eq.\eqref{delta} $u_i$ should be understood as the function $\pi^\ast u_i$ on $\cx$.}
\be\label{delta}
\mathscr{L}_\ce\, u_i=\Delta_i\,u_i
\ee 
The $r$-tupe of numbers $\{\Delta_1,\cdots,\Delta_r\}$ is called the \emph{Coulomb dimensions}.
If the geometry describes an unitary $\cn=2$ SCFT without free subsectors, the $\Delta_i$'s are rational numbers
$>1$. In any given rank $r$ only a finite list of $r$-tuples $\{\Delta_i\}$ are consistent with the geometric structures underlying a special geometry
\cite{caorsi}. The holomorphic differential
\be
\lambda\overset{\rm def}{=} \iota_\ce\Omega
\ee
is called the Seiberg-Witten differential.

Let $\cd\subset \C^r$ be the subset of points with a singular fiber. It is a closed analytic subset of pure codimension 1 called the
\emph{(reduced)  discriminant}. We write $\cd_i$ for its irreducible components and write $\cd=\sum_i\cd_i$ (as divisors in $\C^r$). 

The fiber $\cx_u$ over a ``good'' point $u\in \C^r\setminus\cd$ is a polarized Abelian variety. For simplicity we assume the
polarization to be \emph{principal}, although this is not really necessary. 
Then we have a period map 
\be
p\colon \C^r\setminus \cd\to \boldsymbol{M}_r,
\ee 
where 
\be
\boldsymbol{M}_r\overset{\rm def}{=} Sp(2r,\Z)\backslash Sp(2r,\R)/U(r)
\ee
is the moduli space of principally polarized Abelian varieties of dimension $r$ (the Siegel variety). The map $p$ sends a ``good'' point $u$ of the Coulomb branch to
the \emph{isoclass} of its Abelian fiber $\cx_u$.

Consider the automorphism $\exp(t \ce)\colon \cx \to \cx$ with $t\in\C$. Identifying
the Coulomb branch with the image of $s$ (preserved by the automorphism), 
we get its action on the Coulomb branch $\C^r$ 
\be
\exp(t\ce)\colon\big(u_1,\cdots, u_r\big)\mapsto \big(e^{t\Delta_1}u_1,\cdots,e^{t\Delta_r}u_r\big).
\ee
It is convenient to write \cite{char}
\be
\{\Delta_1,\cdots,\Delta_r\}= \lambda\{d_1,\cdots, d_r\},
\ee
where $\{d_1,\cdots, d_r\}$ are the unique positive integers with $\gcd(d_i)=1$ which represent the point
$\{\Delta_1,\cdots, \Delta_r\}\in \mathbb{P}^{r-1}(\mathbb{Q})$. This shows that the group acting effectively on $\C^r$
is $\C^\times$
\be
\big(u_1,\cdots, u_r\big)\mapsto \big(\zeta^{d_1}u_1,\cdots,\zeta^{d_r}u_r\big),\qquad \zeta\equiv e^{t\lambda}\in\C^\times.
\ee
Each connected component $\cd_i$ of the discriminant is preserved by this $\C^\times$ action,
i.e. $\cd_i$ is the closure of the union of all the $\C^\times$-orbits it contains.
The origin $0\in \C^r$ -- the unique point in the Coulomb branch where the superconformal symmetry is not
spontaneously broken and also the unique closed $\C^\times$-orbit -- then belongs to $\cd_i$ for all $i$. 

\subparagraph{The projective Abelian family $\mathscr{F}$.}
Let $u\in\C^r\setminus \cd$ be a ``good'' point. Since $\exp(t \ce)$ is an automorphism,
the fibers $\cx_u$ and $\cx_{\exp(t\ce)u}$ are isomorphic as polarized Abelian varieties,
i.e. the period map $p$ takes the same value on all points of each $\C^\times$-orbit in $\C^r$.
Said differently: the period map $p$ factors through the projective period map $\varpi$
\be
\varpi\colon (\C^r\setminus \cd)/\C^\times\to \boldsymbol{M}_r.
\ee       
Now
\be
(\C^r\setminus \cd)/\C^\times =\mathbb{P}(d_1,\cdots, d_r)\setminus S_\cd
\ee 
where $\mathbb{P}(d_1,\cdots, d_r)$ is the weighted projective space of weights $(d_1,\cdots,d_r)$
\cite{dolga} and $S_\cd$ is the weighted projective hypersurface whose quasi-cone is the discriminant $\cd$ \cite{dolga}.

In other words, any scale invariant special geometry with given Coulomb dimensions $\{\Delta_i\}$
defines a family $\mathscr{F}$ of (principally) polarized Abelian varieties of dimension $r$ parametrized by
the complement of a hypersurface $S$ in the $(r-1)$-dimensional weighted projective space $\mathbb{P}(d_1,\cdots, d_r)$.
We claim that we have a dichotomy:
\begin{itemize}
\item either the family $\mathscr{F}$ is isotrivial, i.e. $\varpi$ is the constant map;
\item or the family $\mathscr{F}$ is rigid in the sense of Faltings-Peters \cite{rig1,rig2}.
\end{itemize}
The validity of the claim for ranks $r\leq 7$ is obvious \cite{rig3}, and since we are mainly interested in the
$r=2$ case, we shall not pursue this point any further.

\bigskip

We may ask the inverse question: given a rigid family $\mathscr{F}$ of polarized Abelian $r$-varieties parametrized by the complement
of a hypersurface $S$ in $\mathbb{P}(d_1,\cdots, d_r)$,
can we find a special geometry over $\C^r$ which corresponds to this family?

Usually the answer is NO. We can, of course, pull back the family $\mathscr{F}$ to a family parametrized by
the ``good locus'' $\C^r\setminus \cd$, but then we must extend its total space over the discriminant locus,
while requiring the resulting space to have a regular holomorphic symplectic structure $\Omega$.
The last requirement is very strong and ``almost all'' families $\mathscr{F}$ do not admit
such a symplectic extension. The existence and regularity of $\Omega$ are quite formidable constraints.

On the other hand, in the rare situation where our family $\mathscr{F}$ \emph{does have} a symplectic
extension to a special geometry, we expect this extension to be \emph{unique.} 
This is the typical situation, and very plausibly it is true in general.

Assuming uniqueness of the symplectic extension, we get the following useful result:
$$\boldsymbol{(\ast)}\quad\text{\begin{minipage}{390pt}Assume that we have two scale-invariant special geometries with the same $r$-tuple
$\{d_1,\cdots,d_r\}$ and suppose that their respective projective families $\mathscr{F}_1$
and $\mathscr{F}_2$ are isomorphic as holomorphic families of polarized Abelian varieties.
Then their parent scale-invariant special geometries are also isomorphic.\end{minipage}} 
$$
  
\section{The $\{2,2\}$ geometry}

The authors of \cite{Argyres:2022lah} exploit the fact that an  Abelian variety of dimension 2 is either the Jacobian of a smooth
hyperelliptic curve of genus 2 or the product of two elliptic curves. They replace the Abelian  
fibration $\pi\colon \cx\to \C^2$ by a corresponding fibration $\check{\pi}\colon\cy\to \C^2$ whose smooth fibers over $\C^2\setminus\cd$
 are genus 2 hyperelliptic curves. The original fibration $\pi\colon \cx\to \C^2$ is then recovered as the Jacobian fibration of $\check{\pi}\colon\cy\to \C^2$.
 Concretely, they write explicit hyperelliptic equations of degree 6 or 5
 with coefficients which are polynomials in the Coulomb branch coordinates $u,v$.
 For instance, for dimensions $\{2,2\}$ their equation is \cite{Argyres:2022lah}
 \be\label{curve}
 y^2 = (ux-v)(x^5 +\tau_1x^3 +\tau_2x^2 +\tau_3x+\tau_4)
 \ee
 where $\tau_1,\tau_2,\tau_3,\tau_4$ are free complex parameters. 
 By construction the hyperelliptic curve over the point $(u,v)\in \C^2\setminus\cd$
 is isomorphic to the one over $(\lambda u, \lambda v)$, so that we get a family $\cg$
 of genus $2$ curves over $\mathbb{P}^1\setminus S_\cd$ whose Jacobian fibration is our
 family $\mathscr{F}$. The symplectic form may be shown to be
 \be
 \Omega=du\wedge \frac{x\,dx}{y}+dv\wedge \frac{dx}{y}.
 \ee
 \medskip
 
 Equation \eqref{curve} leads to a puzzle. Physically we expect this geometry to correspond to the $\cn=2$ theory of class
 $\cs[A_1]$ with Gaiotto curve the sphere with five punctures \cite{Gaiotto:2009we}, whose conformal manifold should be
 $\cm_{0,5}$, the moduli space of the 5-punctured sphere, whose complex dimension is 2 not 4.
 Why 2 is easy to understand: physically the marginal deformations should be the elements of $\C[u,v]$
 of scaling dimension 2, and there are only two of them for dimensions $\{2,2\}$.

 How we reconcile these physical facts
   with the findings in \cite{Argyres:2022lah} that the geometry \eqref{curve} depends on \emph{four} parameters?
 
 \medskip
 
 \subparagraph{Solving the puzzle.}The crucial ingredient is eq.(A.31) of \cite{Argyres:2022lah}
  which yields the discriminant for this geometry.
 The \emph{reduced} discriminant $\cd$ is
 \be
 \prod_{i=1}^5(u+\alpha_i(\tau) v)
 \ee
 where $\alpha_i(\tau)$ are certain functions of the four parameters $\tau_1,\tau_2,\tau_3,\tau_4$ (called collectively $\tau$).
For a given value $\tau$ of the parameters, the base of the family $\mathscr{F}(\tau)$  is then
 \be
 B(\tau)=\mathbb{P}^1\setminus\{-1/\alpha_1(\tau),\cdots, -1/\alpha_5(\tau)\}
 \ee
which is a 5-punctured sphere. The monodromy around each puncture may be read from table 1
 of \cite{Argyres:2022lah}; each one of them is conjugated over $Sp(4,\Z)$ to the unipotent element
 \be
 \begin{pmatrix} 1 & 2 &0 &0\\
 0 & 1 & 0 & 0\\
 0 & 0 &1 & 0\\
 0 & 0 & 0 & 1\end{pmatrix}
 \ee
 Thus the family $\mathscr{F}$ may be completed to a semistable fibration $\check{\mathscr{F}}\to \mathbb{P}^1$; the Seiberg-Witten differential
yields a holomorphic contact structure on its total space.

 In particular the geometry is non-isotrivial, hence rigid.\footnote{\ This also follows from the fact that 
 the family $\mathscr{G}$ is rigid by Arakelov theorem \cite{arakelov}.}
By rigidity\footnote{\ Here we are cheating a little bit. We need a slightly stronger notion of rigidity: what we are really using here is that the periods of the Seiberg-Witten differential, i.e. the \emph{multivalued} special coordinates of Special Geometry, have the form $\sqrt{u}\, f(v/u)$
where the functions $f(z)$ are \emph{Pochhammer transcendents}.}  
  the family $\mathscr{F}(\tau)$ is uniquely defined by the five special points $-1/\alpha_i(\tau)\in\mathbb{P}^1$
  and the conjugacy classes of the local monodromies around them.
  Since the monodromy classes are fixed, two families $\mathscr{F}(\tau^\prime)$ and $\mathscr{F}(\tau)$
 are isomorphic if and only if
  $B(\tau^\prime)\simeq B(\tau)$ as 5-punctured spheres, i.e. iff they correspond to the same point
  in $\cm_{0,5}$, that is, iff the two sets  of five points $\{\alpha_i(\tau^\prime)\}$ and $\{\alpha_i(\tau)\}$
  have the same cross-ratios.
  
We conclude that the inequivalent families are parametrized by $\cm_{0,5}$.
Using the remark $\boldsymbol{(\ast)}$ at the end of the previous section, we conclude that:
  
  \textit{the non-isomorphic scale-invariant special geometries with dimensions $\{2,2\}$
  are parametrized by the two-dimensional complex space $\cm_{0,5}$.}
  
Up to isomorphism, the special geometry does not depend on the four parameters $\tau_a$ individually,
but only on two cross-ratios of their functions $\alpha_i(\tau)$.
  
  \medskip
  
  We stress that the above result is precisely Gaiotto's prediction \cite{Gaiotto:2009we} for the conformal manifold of 
  the  $SU(2) \times SU(2)$ superconformal gauge theory with a bi-fundamental hypermultiplet as well as two fundamental hypermultiplets for each gauge factor.

\end{document}